\documentclass[review]{elsarticle}

\usepackage{lineno,hyperref}
\modulolinenumbers[5]

\usepackage{longtable}
\usepackage{booktabs}

\journal{Journal of \LaTeX\ Templates}

\begin{document}

\begin{frontmatter}

\title{Graph-based Deep Learning for Communication Networks: A Survey}

\author{Weiwei Jiang\corref{mycorrespondingauthor}}
\address{Department of Electronic Engineering, Tsinghua University, Beijing 100084, China}
\cortext[mycorrespondingauthor]{Corresponding author. E-mail address: jwwthu@gmail.com}

\begin{abstract}
Communication networks are important infrastructures in contemporary society. There are still many challenges that are not fully solved and new solutions are proposed continuously in this active research area. In recent years, to model the network topology, graph-based deep learning has achieved the state-of-the-art performance in a series of problems in communication networks. In this survey, we review the rapidly growing body of research using different graph-based deep learning models, e.g. graph convolutional and graph attention networks, in various problems from different types of communication networks, e.g. wireless networks, wired networks, and software defined networks. We also present a well-organized list of the problem and solution for each study and identify future research directions. To the best of our knowledge, this paper is the first survey that focuses on the application of graph-based deep learning methods in communication networks involving both wired and wireless scenarios. To track the follow-up research, a public GitHub repository is created, where the relevant papers will be updated continuously.
\end{abstract}


\end{frontmatter}


\section{Introduction}
\label{sec:intro}
Communication networks are ubiquitous in contemporary society, from the widely used Internet and 4G/5G cellular networks to the fast-growing Internet of Things (IoT) networks. The growing of communication networks has gone beyond the imagination of their designers. For example, based on Cisco Annual Internet Report (2018–2023) White Paper, nearly two-thirds of the global population will have Internet access by 2023~\footnote{\url{https://www.cisco.com/c/en/us/solutions/collateral/executive-perspectives/annual-internet-report/white-paper-c11-741490.html}}. It would be very challenging to operate and manage such giant networks and new network types keep bringing new problems. For example, the manual configuration becomes infeasible or inefficient in modern networks. While the research for communication networks has a long history, it is still an active area with a steady stream of new ideas, e.g., Software Defined Networking (SDN) and Space-Air-Ground Integrated Network (SAGIN). The challenges may not only include the traditional ones, e.g., routing and load balancing, power control and resource allocation, but also the emerging ones, e.g., virtual network embedding in SDN.

To solve these challenges, various solutions are introduced to the networking domain, especially deep learning~\cite{goodfellow2016deep}. Represented by deep neural networks, deep learning has achieved a great success in many problems, especially in image recognition, natural language processing, and time series problems~\cite{he2016deep, jiang2018geospatial, young2018recent, jiang2021applications, jiang2020time}. Deep learning models are also applied in various communication networks and are proven extremely useful for a series of problems, e.g., network design, traffic prediction, resource allocation, etc~\cite{zappone2019wireless, zhang2019deep, wang2020thirty, abbasi2021deep}. However, in these studies, the network topology structure is not fully utilized because most of the deep neural networks are designed for Euclidean structure data, e.g., images and videos. To amend this shortcoming, graph-based deep learning represented by Graph Neural Networks (GNNs) are proposed for non-Euclidean structure data in recent years~\cite{wu2020comprehensive, zhou2020graph, zhang2020deeplearning, xia2021graph, ruiz2021graph, jiang2021graph}. More recently, GNNs are combined with deep reinforcement learning for making decisions in a series of problems, e.g., GNN is used for processing the graph information and  improving the inter-coflow scheduling ability in distributed computing~\cite{sun2021deepweave}.

GNNs are suitable for problems in communication networks because of their strong learning ability to capture the spatial information hidden in the network topology and their generalization ability to be used in unseen topologies when the networks are dynamic. As to be discussed in this survey, GNN-based solutions are proven effective for a wide range of problems in different network scenarios and are worthy of being explored deeper in the future.

To the best of the authors' knowledge, this paper presents the first literature survey of graph-based deep learning studies for problems in communication networks, covering a total of 81 papers ranging from 2016 to 2021 and involving both wired and wireless scenarios. Compared to a recent similar survey~\cite{he2021overview} which only covers the applications of GNNs in wireless networks, our survey has a broader coverage and contains almost all the surveyed studies from~\cite{he2021overview}. The scope of communication networks used in this survey is broad, thus the surveyed papers are selected from a wide range of journals and conferences. Because it is still a very rapidly developing research topic of applying graph-based deep learning methods, we also include preprints that have not yet gone through the traditional peer review process (e.g., arXiv papers) to present the latest progress. 

\begin{figure}[!htb]
    \centering
    \includegraphics[width=\textwidth]{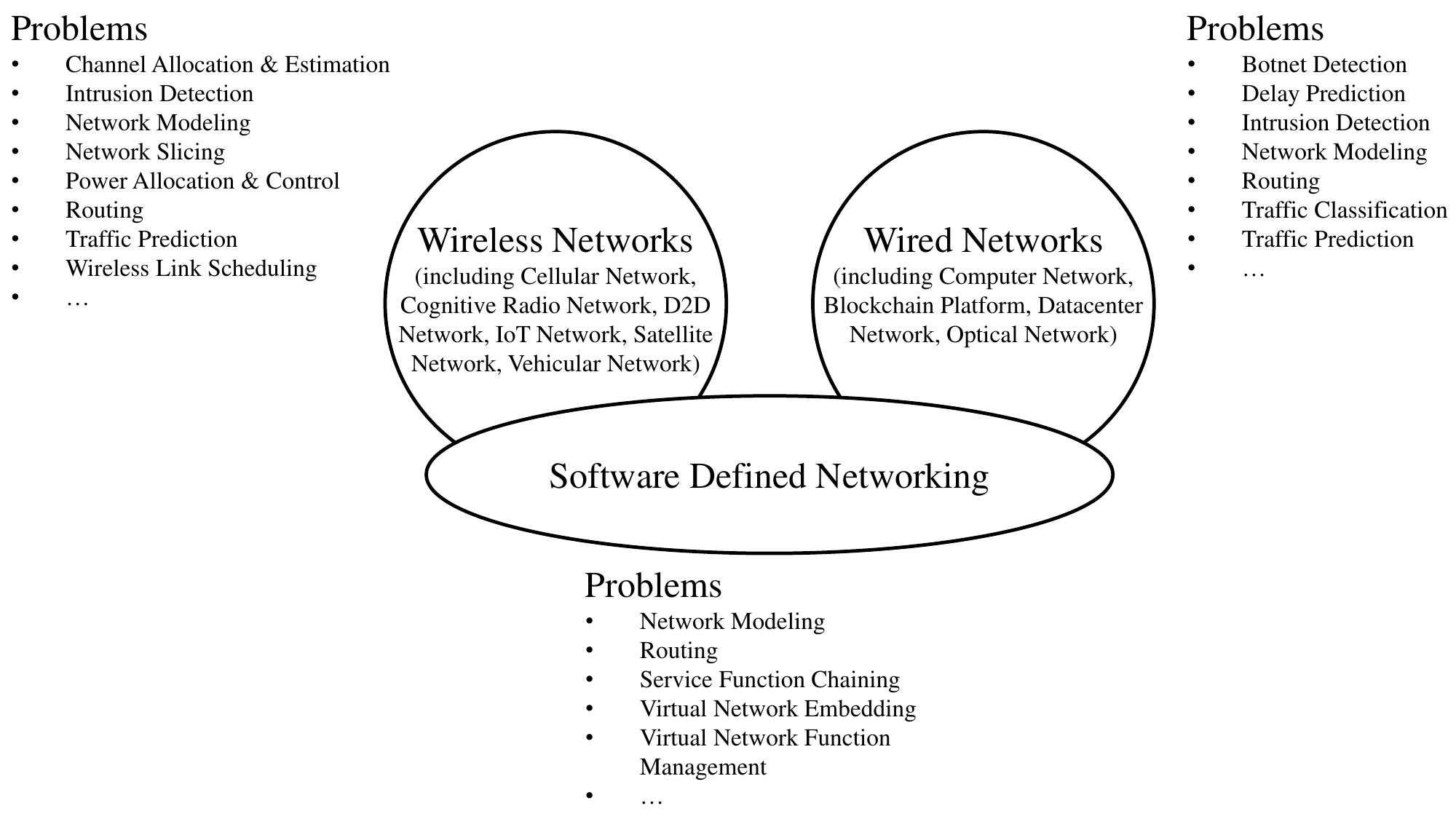}
    \caption{The organization of this survey.}
    \label{fig:fig1}
\end{figure}

The surveyed papers are classified into three major scenarios, as organized in Figure~\ref{fig:fig1}. Some of the common problems are discussed in two or three scenarios, e.g., network modeling, routing, traffic prediction. The other problems are only mentioned in one of these scenarios. This kind of organization is not exclusive, because the idea of SDN can be applied for both the wireless and wired networks. Graph-based deep learning is being frequently used in the assumption of future softwarized networks, without a strict constraint about which type of substrate network is being used. By taking the SDN scenario as a separate section, the relevant discussion would be inspiring for both the future work in the wireless and wired scenarios.

In this survey, the problems to solve, the graph-based solutions and the specific GNN models used in each study are identified and summarized. We also attempt to point out the future directions of applying GNNs in communication networks. Our aim is to provide an up-to-date summary of related work and a useful starting point for new researchers interested in related topics. In addition to this paper, we have also created an open GitHub repository~\footnote{\url{https://github.com/jwwthu/GNN-Communication-Networks}} to update new papers continuously.

Our contributions are summarized as follows:

1) \textit{Comprehensive Review}: We present the up-to-date comprehensive review of graph-based deep learning solutions for problems in various types of communication networks, in the past six years (2016-2021).

2) \textit{Well-organized Summary}: We summarize the problem to solve, the graph-based solution and the GNNs used in each study in a unified format, which would be useful as a reference manual.

3) \textit{Future Directions}: We propose several potential future directions for researchers interested in relevant topics. 

For reference, the list of the acronyms frequently used in this survey is summarized in Table~\ref{tab:acronyms}.

\begin{center}
\begin{longtable}{lp{10cm}}
\caption{The list of the acronyms used in this survey.} \label{tab:acronyms} \\
\hline Acronym & Full Name \\ \hline 
\endfirsthead

\multicolumn{2}{c}%
{{\bfseries \tablename\ \thetable{} -- continued from previous page}} \\
\hline Acronym & Full Name \\ \hline 
\endhead

\hline \multicolumn{2}{r}{{Continued on next page}} \\ \hline
\endfoot

\hline
\endlastfoot

BGP & Border Gateway Protocol \\
DC-STGCN & Dual-Channel based Graph Convolutional Network \\
DCRNN & Diffusion Convolutional Recurrent Neural Network \\
DL & Deep Learning \\
DQN & Deep Q Network \\
DRL & Deep Reinforcement Learning \\
FDS-MARL & Fully Decentralized Soft Multi-Agent Reinforcement Learning \\
GASTN & Graph Attention Spatial-Temporal Network \\
GAT & Graph Attention Network \\
GCLR & GNN based Cross Layer optimization by Routing \\
GCN & Graph Convolutional Network \\
GE & Graph Embedding \\
GGS-NN & Gated Graph Sequence Neural Network \\
GIN & Graph Isomorphism Network \\
GN & Graph Network \\
GNN & Graph Neural Network \\
HIGNN & Heterogeneous Interference Graph Neural Network \\
HetGAT & Heterogeneous Graph Attention Network \\
IGCNet & Interference Graph Convolutional Neural Network \\
ML & Machine Learning \\
MPGNNs & Message Passing Graph Neural Networks \\
MPLS & Multiprotocol Label Switching \\
MPNN & Message Passing Neural Network \\
MSTNN & Multi-scale Spatial-Temporal Graph Neural Network \\
NFV & Network Function Virtualization \\
REGNNs & Random Edge Graph Neural Networks \\
S-RNN & Structural-RNN \\
SDN & Software Defined Networking \\
SFC & Service Function Chaining \\
SGCRN & Spatiotemporal Graph Convolutional Recurrent Network \\
TCN & Temporal Convolutional Network \\
TGCN & Temporal Graph Convolutional Network \\
UWMMSE & Unfolded iterative Weighted Minimum Mean Squared Error \\
VNE & Virtual Network Embedding \\
VNF & Virtual Network Function \\
\hline

\end{longtable}
\end{center}

The remainder of this paper is organized as follows. In Section~\ref{sec:method}, we introduce the progress of conducting literature search and selection. In Section~\ref{sec:gnns}, we introduce the GNNs used in the reviewed studies. In Section~\ref{sec:wireless}, we summarize the studies in wireless networks. In Section~\ref{sec:wired}, we summarize the studies in wired networks. In Section~\ref{sec:sdn}, we summarize the studies in software defined networks. In Section~\ref{sec:direction}, we point out future directions. In Section~\ref{sec:conclusion}, we draw the conclusion.

\section{Survey Methodology}
\label{sec:method}
To collect relevant studies, the literature is searched with various combinations of two groups of keywords. The first group is about the graph-based deep learning techniques, e.g., ``Graph", ``Graph Embedding", ``Graph Neural Network", ``Graph Convolutional Network", ``Graph Attention Networks", ``GraphSAGE", ``Message Passing Neural Network", ``Graph Isomorphism Network", etc. The second group is about the communication networks as well as specific problems, e.g., ``Wireless Network", ``Cellular Network", ``Computer Network", ``Software Defined Networking", ``Traffic Prediction", ``Routing", ``Service Function Chaining", ``Virtual Network Function", etc. The databases from major publishers are carefully covered one by one, e.g., ACM, Elsevier, IEEE, Springer, Wiley, etc. To track the citation relationship among these papers and avoid missing records from smaller publishers, Google Scholar is also leveraged in the literature search process.

\begin{figure}[!htb]
    \centering
    \includegraphics[width=\textwidth]{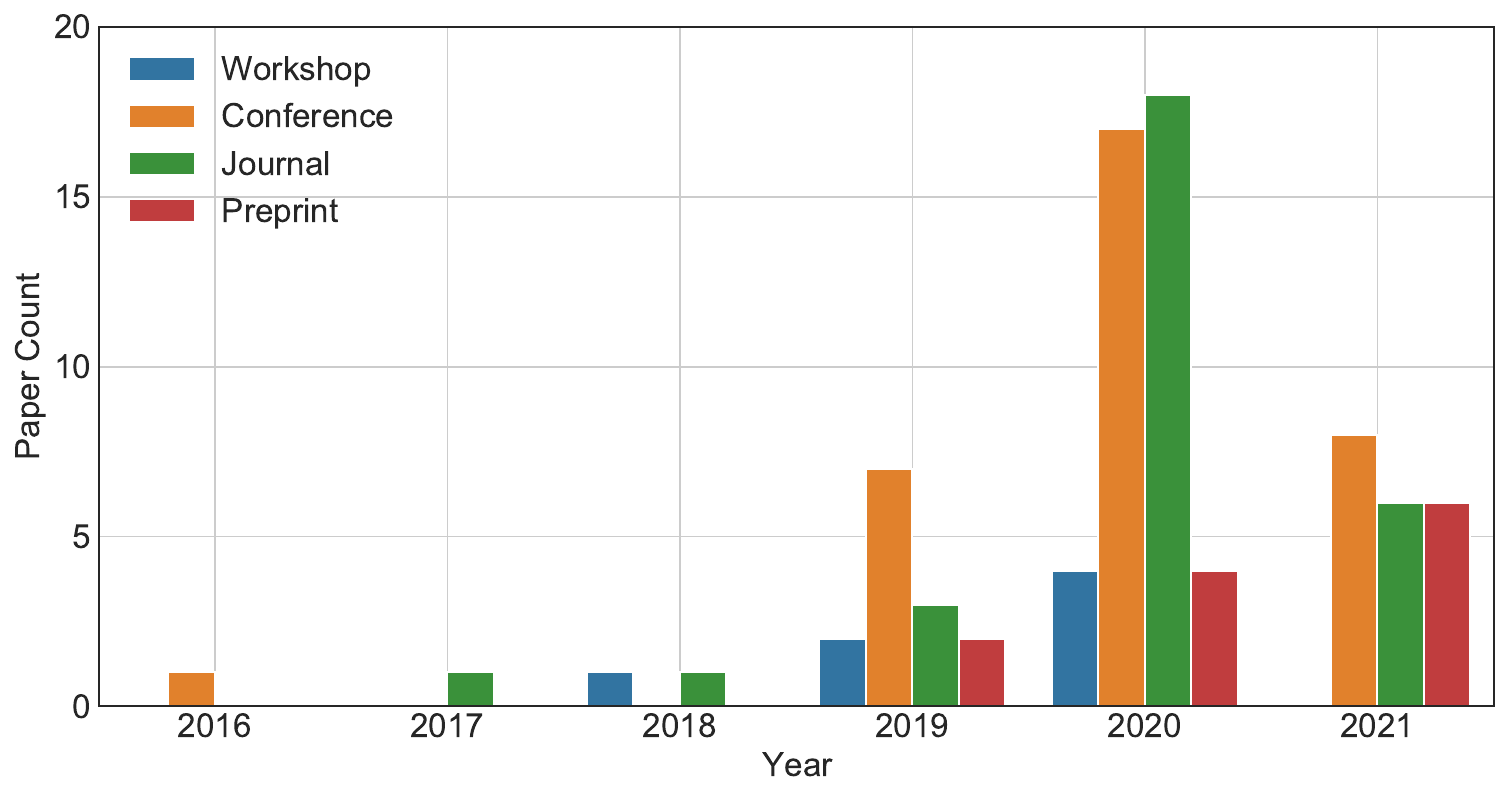}
    \caption{The paper count of different types annually.}
    \label{fig:paper_count}
\end{figure}

\begin{figure}[!htb]
    \centering
    \includegraphics[width=\textwidth]{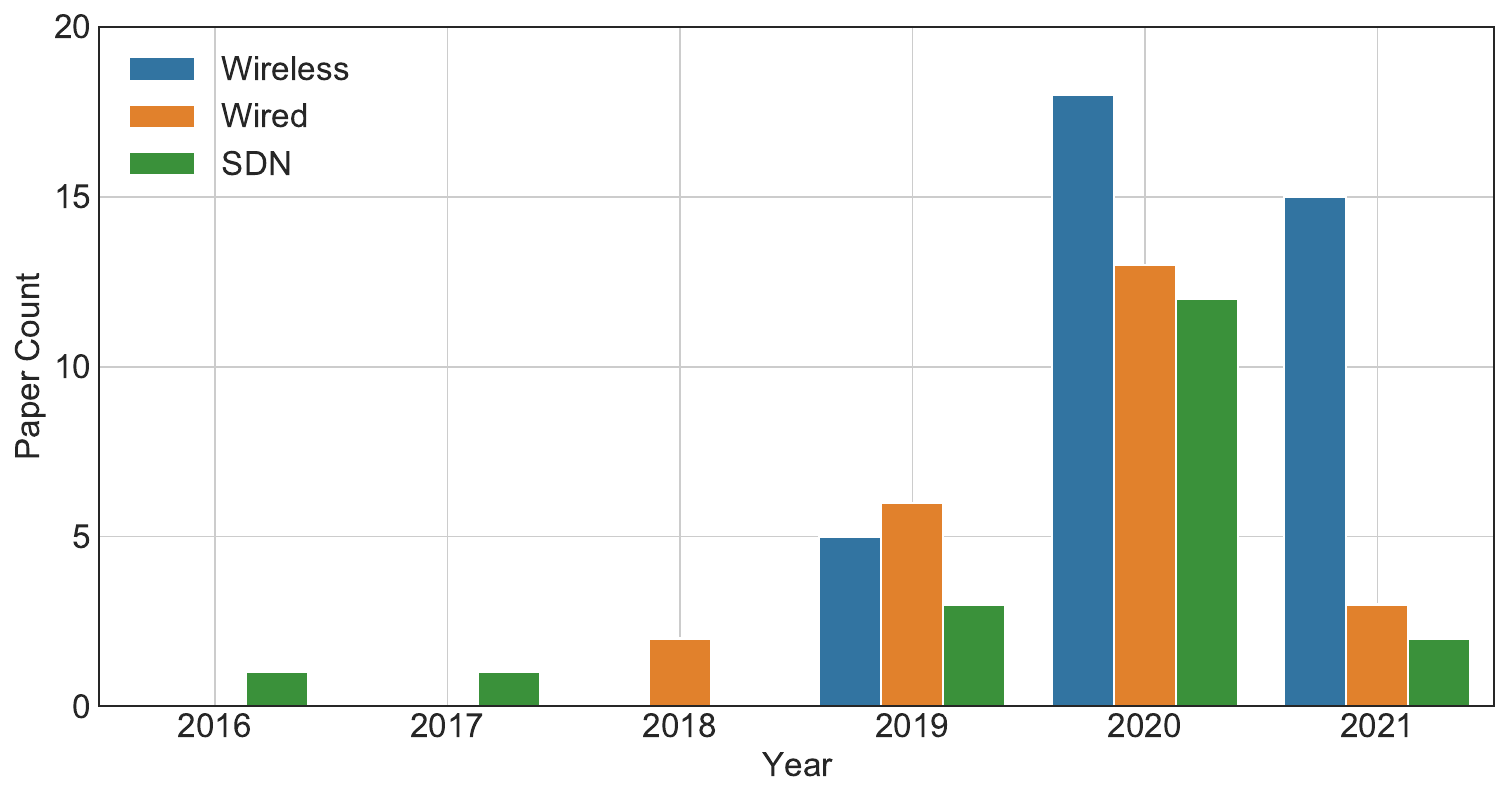}
    \caption{The paper count of different network types annually.}
    \label{fig:type_count}
\end{figure}

A total of 81 papers are finally selected and covered in this survey, with the earliest one published in year 2016, as shown in Figure~\ref{fig:paper_count}. Most of the surveyed papers are published in recent three years, i.e., 2019, 2020, and the first five months of 2021. Compared with 14 papers in 2019, there is a 207\% growth of papers in 2020, with a total of 43 papers. While there are only 20 papers in the first five months of 2021, it is expected that more relevant studies would be published or publicized in the remaining months with the growing impact of graph-based deep learning methods being applied in the networking domain. We also show the paper statistics for different network types in Figure~\ref{fig:type_count}. The wireless network scenario draws more attention than the other two and this trend may continue in 2021.

For a full coverage of relevant studies, workshop, conference, and journal papers as well as preprint papers are covered in this survey, to track the latest achievements as well as the on-going progress. The journal list (alphabetically) is shown in Table~\ref{tab:journal}. The conference list (alphabetically) is shown in Table~\ref{tab:conference}. And the workshop list (alphabetically) is shown in Table~\ref{tab:workshop}. All the preprint papers are from the arXiv platform~\footnote{\url{https://arxiv.org/}}. Since we cover a wide area with various communication networks, the papers are selected from various publications or conference proceedings, some of which may focus on telecommunications or related subjects and the others may be multidisciplinary. As an emerging topic which has not been widely adopted, graph-based deep learning appears in recent years for solving networking-related problems, with only one paper selected for most journals or conferences.

\begin{table}[!htb]
\centering
\caption{List of source journals and the corresponding studies we cover in this study.}
\label{tab:journal}
\begin{tabular}{ll}
\hline
Journal Name & Studies \\
\hline
Computer Networks & \cite{sun2020efficient, li2020traffic} \\
Electronics & \cite{pan2021dc} \\
IEEE Access & \cite{nakashima2020deep, zhu2020gclr} \\
IEEE Communications Letters & \cite{cheng2021discovering, sun2020combining, simsek2020iab, rusek2018message} \\
IEEE Internet of Things Journal & \cite{liu2020dynamic} \\
IEEE Journal on Selected Areas in Communications & \cite{fang2019idle, rusek2020routenet, yan2020automatic, shen2020graph} \\
IEEE Systems Journal & \cite{zhuang2019toward} \\
IEEE Transactions on Industrial Informatics & \cite{wang2020graph} \\
IEEE Transactions on Information Forensics and Security & \cite{shen2021accurate} \\
IEEE Transactions on Mobile Computing & \cite{sun2021mobile, he2020graph} \\
IEEE Transactions on Network Science and Engineering & \cite{geyer2020graph} \\
IEEE Transactions on Network and Service Management & \cite{mijumbi2017topology} \\
IEEE Transactions on Signal Processing & \cite{eisen2020optimal} \\
IEEE Transactions on Vehicular Technology & \cite{yan2020cooperative} \\
IEEE Transactions on Wireless Communications & \cite{chowdhury2021unfolding, lee2020graph} \\
International Journal of Network Management & \cite{kim2021graph} \\
Performance Evaluation & \cite{geyer2019deepcomnet} \\
Sensors & \cite{zhao2020graph} \\
Transactions on Emerging Telecommunications Technologies & \cite{zhao2020spatiotemporal} \\
\hline
\end{tabular}
\end{table}

\begin{table}[!htb]
\centering
\caption{List of source conferences and the corresponding studies we cover in this study.}
\label{tab:conference}
\begin{tabular}{p{12cm}l}
\hline
Conference Name & Studies \\
\hline
ACM SIGCOMM conference & \cite{suarez2019challenging} \\
ACM Symposium on SDN Research (SOSR) & \cite{rusek2019unveiling} \\
Asia-Pacific Network Operations and Management Symposium (APNOMS) & \cite{kim2020graph, heo2020graph} \\
IEEE Annual Consumer Communications \& Networking Conference (CCNC) & \cite{rkhami2021learn} \\
IEEE Conference on Computer Communications (INFOCOM) & \cite{geyer2019deeptma} \\
IEEE Conference on Network Function Virtualization and Software Defined Networks (NFV-SDN) & \cite{jalodia2019deep} \\
IEEE Global Communications Conference (GLOBECOM) & \cite{he2020resource, he2019graph} \\
IEEE International Conference on Acoustics, Speech and Signal Processing (ICASSP) & \cite{chowdhury2021efficient, eisen2020transferable} \\
IEEE International Conference on Communications (ICC) & \cite{yang2020mstnn,sun2020deepmigration, lee2020wireless, tekbiyik2021channel, wang2021drl} \\
IEEE Symposium on Computers and Communications (ISCC) & \cite{geyer2020robustness} \\
IEEE Vehicular Technology Conference (VTC) & \cite{fu2020wireless} \\
IEEE Wireless Communications and Networking Conference (WCNC) & \cite{guo2021learning, shao2021graph, hou2021user} \\
IFIP Networking Conference (IFIP Networking) & \cite{geyer2019deepmpls} \\
International Conference on Information Networking (ICOIN) & \cite{sawada2020network, suzuki2020estimating} \\
International Conference on Information and Communication Technology Convergence (ICTC) & \cite{rafiq2020service} \\
International Conference on Network and Service Management (CNSM) & \cite{kim2020graph1, habibi2020accelerating, mijumbi2016connectionist} \\
International Conference on Real-Time Networks and Systems (RTNS) & \cite{mai2021improvements} \\
International Conference on Wireless Communications and Signal Processing (WCSP) & \cite{yang2020noval} \\
International Conference on emerging Networking EXperiments and Technologies (CoNEXT) & \cite{badia2019towards} \\
International Symposium on Networks, Computers and Communications (ISNCC) & \cite{rkhami2020use} \\
Opto-Electronics and Communications Conference (OECC) & \cite{gui2020optical} \\
\hline
\end{tabular}
\end{table}

\begin{table}[!htb]
\centering
\caption{List of source workshops and the corresponding studies we cover in this study.}
\label{tab:workshop}
\begin{tabular}{p{12cm}l}
\hline
Workshop Name & Studies \\
\hline
AutoML for Networking and Systems Workshop of MLSys Conference & \cite{zhou2020auto} \\
IEEE Globecom Workshops (GC Wkshps) & \cite{shen2019graph} \\
IEEE International Workshop on Signal Processing Advances in Wireless Communications (SPAWC) & \cite{eisen2019large, naderializadeh2020wireless} \\
Workshop on Big Data Analytics and Machine Learning for Data Communication Networks & \cite{geyer2018learning} \\
Workshop on Network Meets AI \& ML & \cite{bahnasy2020deepbgp, xiao2020neural} \\
\hline
\end{tabular}
\end{table}

\section{Graph-based Deep Learning Introduction}
\label{sec:gnns}
In this section, we first present some typical examples of the graph structures used in communication networks. Then we give a short introduction of the graph-based deep learning models, especially those used in the surveyed papers. Finally, we discuss the pros and cons of applying graph-based deep learning models in the networking domain.

\subsection{Graphs in Communication Networks}
From the graph theory, a simple graph is defined as $G=(V, E)$, where $V$ is the set of nodes and $E$ is the set of edges between nodes. In communication networks, the edges can be either directed or undirected, depending on the specific problems. Both nodes and edges can be associated with some attributes as the features, either static or dynamic.

\begin{figure}[!htb]
    \centering
    \includegraphics[width=\textwidth]{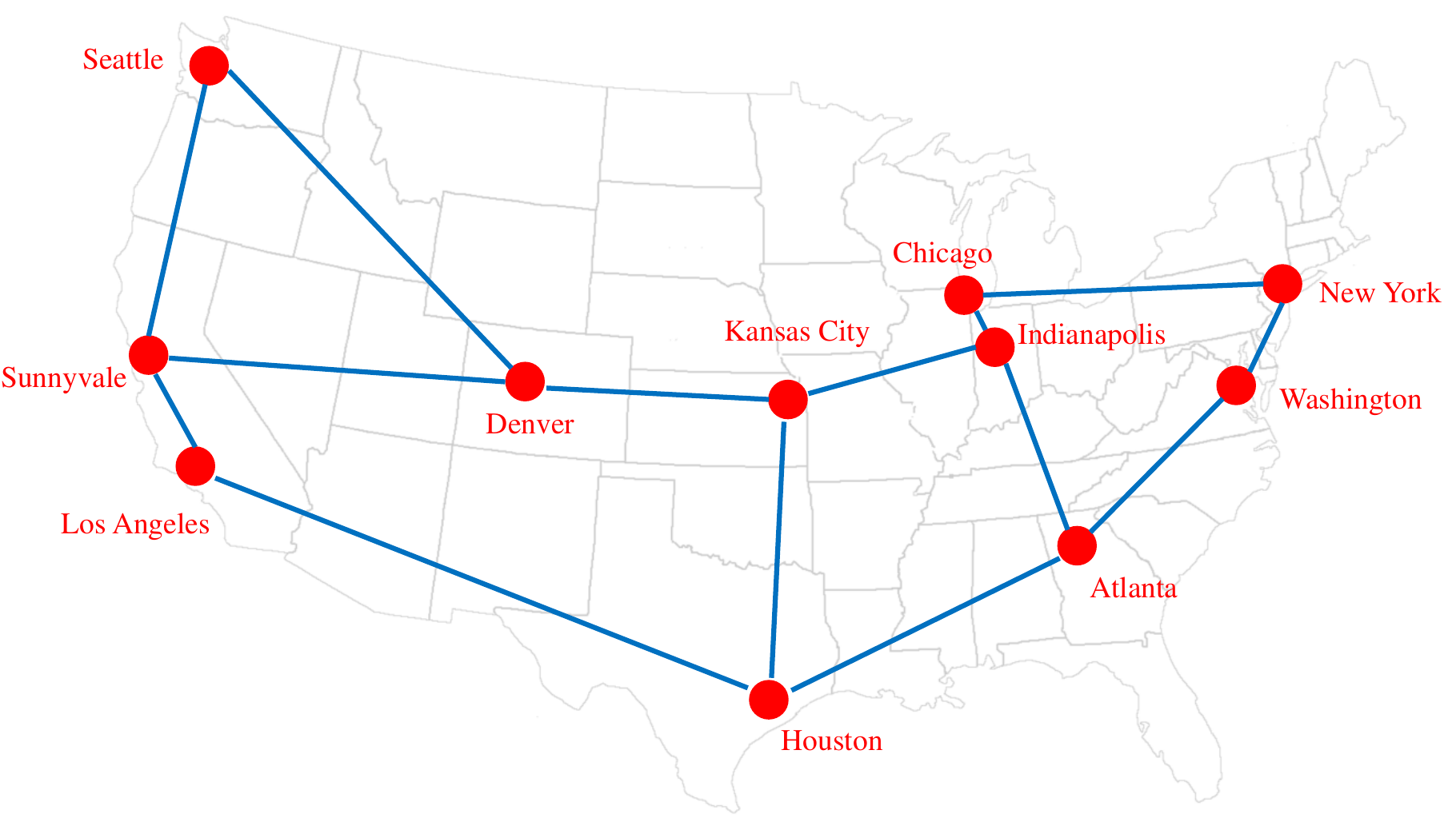}
    \caption{An example of the communication graph from the Abilene network.}
    \label{fig:fig3}
\end{figure}

Two graph examples are given for the wired and wireless scenarios respectively. In Figure~\ref{fig:fig3}, the communication graph from the Abilene network is presented, which consists of 11 nodes and 14 edges. Each node represents the physical backbone router and the node features include the inflow and outflow traffic volumes. Each edge represents the physical transmission link and the edge features include the transmission metrics, e.g., bandwidth and delay. Similar communication graphs are built from other network topologies, e.g., the Nobel, G\'{E}ANT, Germany50, and AT\&T backbone networks, can be found in~\cite{zhao2020spatiotemporal, yang2020mstnn, zhuang2019toward}.

\begin{figure}[!htb]
    \centering
    \includegraphics[width=\textwidth]{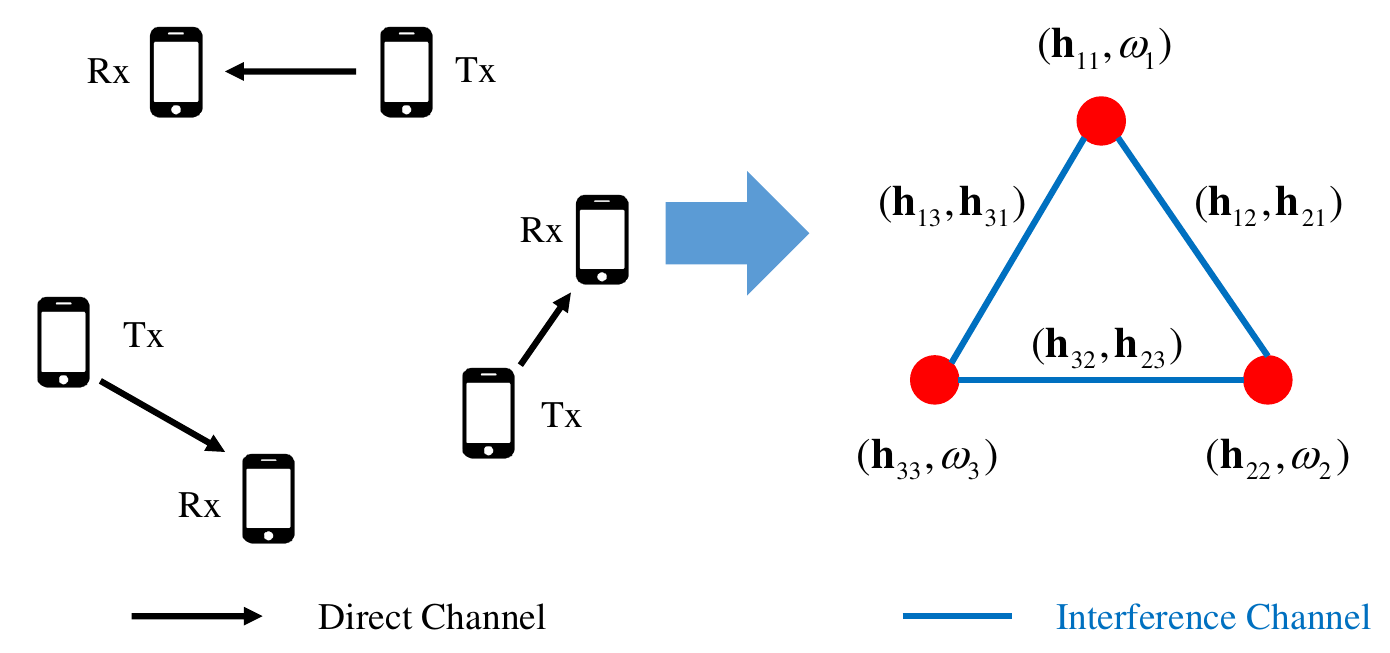}
    \caption{An example of the interference graph from~\cite{shen2019graph}.}
    \label{fig:fig4}
\end{figure}

In Figure~\ref{fig:fig4}, the interference graph for a homogeneous ad-hoc network is presented, which consists of 3 nodes and 3 edges. Different from Figure~\ref{fig:fig3}, the nodes in Figure~\ref{fig:fig4} are virtual nodes, each of which corresponds to a transceiver pair (Tx, Rx). The features of node $i$ include the direct channel state information (CSI) $\mathbf{h}_{ii}$ and other environmental information, e.g., the weight $\omega_i$ of node $i$~\cite{shen2019graph}. The undirected edge between node $i$ and node $j$ models the interference between two transceiver pairs and the edge features are the interference CSIs $\mathbf{h}_{ij}$ and $\mathbf{h}_{ji}$. The interference graph built for the heterogeneous ad-hoc network case can be further found in~\cite{zhang2021scalable}. 

An adjacency matrix $\mathbf{A}$ is introduced to incorporate the network topology information into the architecture of neural networks. Let $e_{ij}$ represents the edge between node $v_i$ and node $v_j$. Then the element of the adjacency matrix $A$ is defined as follows: $A_{ij} = 1$ if $e_{ij} \in E$, otherwise, $A_{ij} = 0$. Here the binary matrix $A$ only captures the connection relationship. If $\mathbf{A}$ is symmetric, the graph is undirected, otherwise, the graph is directed. More complex adjacency matrices can be defined similarly, e.g., the distance matrix or the interference matrix. 

For defining the GNNs in the next part, more notations are introduced here. Based on the connection relationship, $\mathcal{N}(v_i)$ represents the neighbor node set of $v_i$ and each element of the degree matrix $\mathbf{D}$ is $\mathbf{D}_{ii}=\|\mathcal{N}(v_i)\|$. The Laplacian matrix of an undirected graph is introduced and defined as $\mathbf{L} = \mathbf{D} - \mathbf{A}$ and the normalized Laplacian matrix is further defined as $\tilde{\mathbf{L}} = \mathbf{I}_N - \mathbf{D}^{-\frac{1}{2}} \mathbf{A} \mathbf{D}^{-\frac{1}{2}}$, where $N$ is the number of nodes and $\mathbf{I}_N$ is the identity matrix with size $N$. The node feature matrix of a graph is defined as $\mathbf{X} \in {R}^{N \times d}$, where $d$ is the dimension of the node feature vector.

\subsection{Graph-based Models in Communication Networks}
\begin{figure}[!htb]
    \centering
    \includegraphics[width=\textwidth]{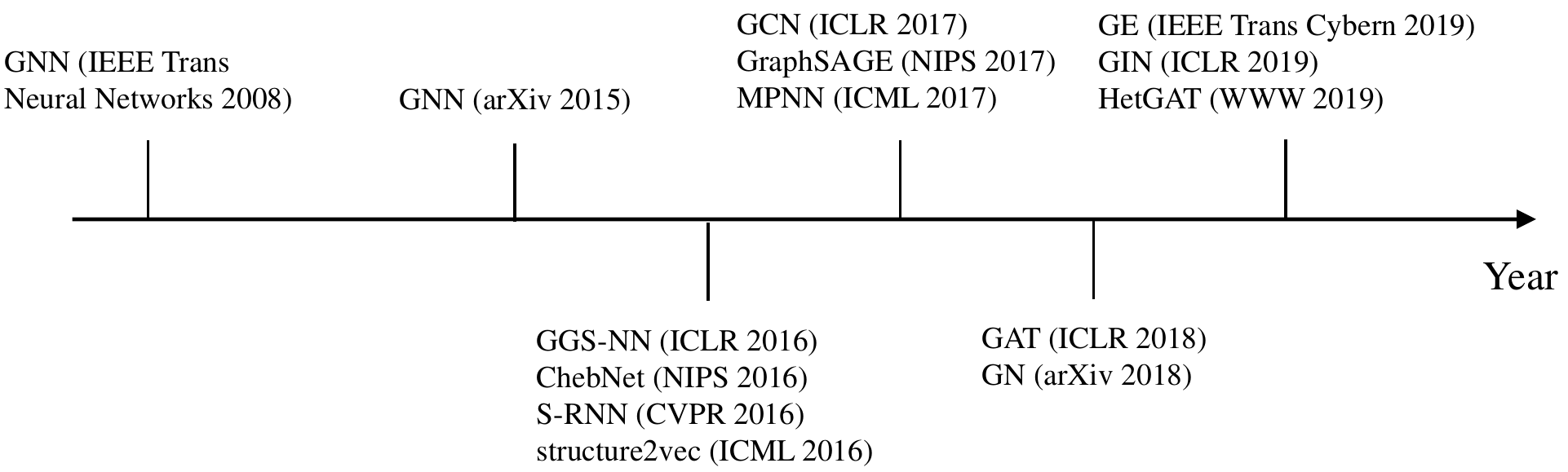}
    \caption{The relevant graph-based deep learning models of this survey.}
    \label{fig:fig2}
\end{figure}

Since the research for graph-based deep learning is still in a fast pace with new models appearing continuously, we have no intention of conducting a thorough literature search on the graph-based models. In this section, we would focus on a short introduction for the GNNs used in the surveyed studies. For those who are interested in the whole picture of graph neural networks and a deeper discussion of the technical details, recent surveys~\cite{wu2020comprehensive, zhou2020graph, zhang2020deeplearning, xia2021graph, ruiz2021graph} are recommended. The relevant graph-based deep learning models are listed chronologically in Figure~\ref{fig:fig2}. Please note that the listed conferences may be lagged behind the preprint versions, which could be released one or two years earlier.

As a pioneering study, GNN is introduced in~\cite{scarselli2008graph}, which extends the application of neural networks from Euclidean structure data to non-Euclidean structure data. GNN is based on the message passing mechanism, in which each node updates its state by exchanging information with each other until it reaches a certain stable state. Afterwards, various GNN variants are proposed, e.g., Graph Convolutional Network (GCN) and Graph Attention Networks (GAT).

We first introduce the Graph Embedding (GE) models. In mathematics, embedding is a mapping function $f: X \rightarrow Y$, in which a point in one space $X$ is mapped to another space $Y$. Embedding is usually performed from a high-dimensional abstract space to a low-dimensional space. Generally speaking, the representation mapped to the low-dimensional space is easier for neural networks to handle with. In the case of graphs, graph embedding is used to transform nodes, edges, and their features into the vector space, while preserving properties like graph structure and information as much as possible. For the studies covered in this survey, several graph embedding models are involved, including structure2vec~\cite{dai2016discriminative}, GraphSAGE~\cite{hamilton2017inductive}, and GE~\cite{pan2019learning}. In a transductive learning approach, Structure2vec~\cite{dai2016discriminative} is based on the idea that if the two sequences composed of all the neighbors of two nodes are similar, then the two nodes are similar. GraphSAGE~\cite{hamilton2017inductive} is a representative of inductive learning. It does not directly learn the representation of each node, but learns the aggregation function instead. For the new node, its embedding representation is generated directly without the need to learn again. Furthermore, a novel adversarial regularized framework is proposed for graph embedding in~\cite{pan2019learning}.

Then we introduce the GCN models. GCN extends the convolution operation from traditional data (such as images) to graph data, inspired by the convolutional neural networks which are extremely successful for image-based tasks. The core idea is to learn a function mapping, through which a node can aggregate its own features and the features of its neighbors to generate the new representation. Generally speaking, there are two types of GCN models, namely, spectral-based and spatial-based. 

Based on graph signal processing, spectral-based GCNs define the convolution operation in the spectral domain, e.g., the Fourier domain. To conduct the convolution operation, a graph signal is transformed to the spectral domain by the graph Fourier transform. Then the result after the convolution is transformed back by the inverse graph Fourier transform. Several spectral-based GCNs are used in the surveyed studies, e.g., GNN~\cite{henaff2015deep}, ChebNet~\cite{defferrard2016convolutional}, and GCN~\cite{kipf2017semi}, which improve the convolution operation with different techniques. By introducing a parameterization with smooth coefficients, GNN~\cite{henaff2015deep} attempts to make the spectral filters spatially localized. ChebNet~\cite{defferrard2016convolutional} learns the diagonal matrix as an approximation of a truncated expansion in terms of Chebyshev polynomials up to $K$th order.

To avoid overfitting, $K=1$ is used in GCN~\cite{kipf2017semi}. More specifically, the graph convolution operation $*G$ in GCN is defined as follows:
\begin{equation}
\mathbf{X}_{*G} = \mathbf{W} (\mathbf{I}_N + \mathbf{D}^{-\frac{1}{2}} \mathbf{A} \mathbf{D}^{-\frac{1}{2}}) \mathbf{X}
\end{equation}
\noindent where $\mathbf{W}$ is a learnable weight matrix, i.e., the model parameters. To alleviate the potential gradient explosion problem, the graph convolution operation is further transformed into:
\begin{equation}
\mathbf{X}_{*G} = \mathbf{W} ( \tilde{\mathbf{D}}^{-\frac{1}{2}} \tilde{\mathbf{A}} \tilde{\mathbf{D}}^{-\frac{1}{2}}) \mathbf{X}
\end{equation}
\noindent where $\tilde{\mathbf{A}} = \mathbf{A} + \mathbf{I}_N$ and $\tilde{\mathbf{D}}_{ii} = \sum_{j}{\tilde{\mathbf{A}}_{ij}}$.

Several spatial-based GCNs are also used in the surveyed studies, which defines the convolution operation directly on the graph based on the graph topology. To unify different spatial-based variants, Message Passing Neural Network (MPNN)~\cite{gilmer2017neural} proposes the usage of message passing functions, which contain a message passing phase and a readout phase. The message passing phase is defined as follows:
\begin{equation}
\mathbf{m}_{v_i}^{(t)} = \sum_{v_j \in \mathcal{N}{(v_i)}} \mathcal{M}^{(t)} (\mathbf{X}_i^{(t-1)}, \mathbf{X}_j^{(t-1)}, \mathbf{e}_{ij})
\end{equation}
\noindent where $\mathbf{m}_{v_i}^{(t)}$ is the message aggregated from the neighbors of node $v_i$, $\mathcal{M}^{(t)}(\cdot)$ is the aggregation function in the $t$-th iteration, $\mathbf{X}_i^{(t)}$ is the hidden state of node $v_i$ in the $t$-th iteration, and $\mathbf{e}_{ij}$ is the edge feature vector between node $v_i$ and node $v_j$. The readout phase is further defined as follows:
\begin{equation}
\mathbf{X}_i^{(t)} = \mathcal{U}^{(t)} (\mathbf{X}_i^{(t-1)},\mathbf{m}_{v_i}^{(t)})
\end{equation}
\noindent where $\mathcal{U}^{(t)}(\cdot)$ is the readout function in the $t$-th iteration.

Graph Network (GN)~\cite{battaglia2018relational} also unifies many GNN variants, by learning node-level, edge-level and graph-level representations. Graph Isomorphism Network (GIN)~\cite{xu2019powerful} takes a step further by pointing out that previous MPNN-based methods are incapable of distinguishing different graph structures based on the graph embedding they produce and adjusting the weight of the central node by a learnable parameter to amend this drawback. Attention-based GNN models can be categorized into the spatial-based type. GAT~\cite{velivckovic2018graph} incorporates the attention mechanism into the propagation step and further utilizes the multi-head attention mechanism to stabilize the learning process, which is defined as follows:
\begin{equation}
\mathbf{X}_i^{(t)} = \|_k \sigma (\sum_{j \in \mathcal{N}{(v_i)}} \alpha^k (\mathbf{X}_i^{(t-1)}, \mathbf{X}_j^{(t-1)}) \mathbf{W}^{(t-1)} \mathbf{X}_j^{(t-1)})
\end{equation}
\noindent where $\|$ is the concatenation operation, $\sigma$ is the activation method, $\alpha^{k}(\cdot)$ is the $k$-th attention mechanism.

Other than the convolution operation, the recurrent operation can also be applied in the propagation module of GNNs. The key difference is that the convolution operations use different weights while the recurrent operations share the same weights. For example, Gated Graph Sequence Neural Network (GGS-NN)~\cite{li2016gated} uses Gated Recurrent Units (GRU) in the propagation step.

In realistic networks, the network topology may change occasionally, e.g., with the addition or deletion of routers, which corresponds to the case of dynamic graphs, instead of static graphs. Several GNN variants are proposed for dealing with dynamic graphs. Diffusion Convolutional Recurrent Neural Network (DCRNN)~\cite{li2018dcrnn_traffic} leverages GNNs to collect the spatial information, which is further used in sequence-to-sequence models. By extending the static graph structure with temporal connections, Structural-RNN (S-RNN)~\cite{jain2016structural} can learn the spatial and temporal information simultaneously.

The last case to discuss is the heterogeneous graph, in which the nodes and edges are multi-typed or multi-modal. For this case, meta-path is introduced as a path scheme which determines the type of node in each position of the path, then one heterogeneous graph can be reduced to several homogeneous graphs to perform graph learning algorithms. To generate the final representation of nodes, graph attention is performed on the meta-path-based neighbors and a semantic attention is used over output embeddings of nodes under all meta-path schemes in Heterogeneous Graph Attention Network (HetGAT)~\cite{wang2019heterogeneous}.

\subsection{Pros and Cons of Graph-based Models}
Machine learning has emerged as a new paradigm to solve various networking problems and to automate network management~\cite{kim2020graph}. Compared with traditional methods, ML models provide many benefits for solving the networking relevant problems. The first advantage is that machine learning models can automatically learn and improve from experiences without being explicitly programmed~\cite{kim2020graph}. Even though it takes some efforts to train a machine learning model, the inference time when applying a trained model is much smaller. These efforts are also inevitable when applying a traditional method based on various optimization techniques which may require a long iteration update process. The second advantage is that the machine learning models are more effective in learning wide and dynamically changing data than statistical and heuristic methods. Based on these advantages, machine learning models, especially the deep learning models, have been widely applied in the networking domain.

The story does not end here. While machine learning has achieved a great success in many research fields, e.g., computer vision, natural language processing and time series processing. Most of these fields use Euclidean domain data, for which the feed forward neural networks, CNNs and RNNs are enough. However, for other fields, e.g., chemistry and biology, these models are inadequate for learning the non-Euclidean graph data, which contain rich relational information between each pair of neighboring elements. Many kinds of graph structure data also exist in the communication networks as introduced earlier, which is beyond the ability of non-GNN machine learning models. Driven by the graph structure data, GNNs are preferable because GNNs can automatically learn a condensed representation of each node in the network that incorporates the information about the node, its neighbors, and their inter-connecting topology~\cite{habibi2020accelerating} and support relational reasoning and combinatorial generalization~\cite{li2020traffic}. 

Besides the ability of handling graph structure, GNNs bring new opportunities for other challenges that have not been fully solved by previous machine learning models, e.g., the complexity in the network state and nonstationarity in networking, with a better generalization ability. Communication networks are complex and dynamic systems, and the overall networking performance may be affected by many factors, e.g., the latency metric affects networking efficiency by defeating network protocols~\cite{zhuang2019toward}. Traditional techniques, e.g., the open shortest path first protocol (OSPF) for routing, are not capable of coping with these challenges. When situations such as link failure and congestion happen, these traditional techniques would not be able to converge quickly with these previously unseen situations. The non-GNN machine learning models will no longer apply when the network topology changes, e.g., link disconnection, and new training data are needed~\cite{li2020traffic}. Since the topology of the network is usually dynamically changed, dynamic graphs are used in GNNs for the actual network. In other words, GNN is able to understand the complex relationship between topology, routing and traffic in networks, and generalizes trained NN parameters over arbitrary topologies, routing schemes and variable traffic intensity~\cite{li2020traffic}. It is also proven that GNNs have a higher training efficiency than other neural networks, for example, GNNs converge $O(n \log n)$ times faster and their generalization error is $O(n)$ times lower theoretically, compared with multilayer perceptrons in a communication networks with $n$ nodes~\cite{shen2021neural}.

Even so, GNNs are not the panacea. There are still some concerns about applying GNNs in the networking domain and not all of them have been fully resolved. The first concern is about the collection of training data for GNNs (and other machine learning models too). Compared with the well-developed research areas with large-scale open benchmark datasets, e.g., ImageNet for computer vision, the training datasets are still rare (at least the open ones) for training the effective GNN models. Even for those already used in the existing studies, the data size is limited and is far from the need of being applied in the actual network.

The second concern is about the depth of GNN models. For other neural networks, e.g., CNNs, it has been proven effective to use a deeper structure, e.g., ResNet. However, similar benefits are not obvious for GNNs. It has been found that when using more than two GCN layers, the performance becomes worse with more GCN layers. This is because GNNs rely on the aggregation operation on the features of neighbor nodes, the results become too smooth and lack of differentiation after multiple layers. As the network continues to overlap, eventually all nodes will learn the same expression and GNNs fail to work. It is still questionable Whether the graph neural network needs a deep structure, or whether a deep network structure can be designed to avoid the problem of over-smoothness in the networking domain.

The third concern is about the stability of GNN models, both under the stochastic perturbations and adversarial attacks~\cite{zugner2018adversarial, gama2020stability, keriven2020convergence}. Stochastic perturbations appear in the communication networks in the situations when link failure and congestion happen. While the adversarial attacks appear when targeted attacks on the underlying networks happen. These problems already exist for other neural networks and more attack types can be designed by leveraging the node features or the graph structure. It has been found that the stability of GNNs is affected by multiple factors, e.g., the graph filter, nonlinearity, architecture width and depth, etc~\cite{gama2020stability}. And massive efforts have been put to design GNNs which are robust to the perturbations or attacks. With the deeper involvement of GNNs with various networking problems, more potential vulnerable cases would appear which would require a design of more robust GNNs.

The last but not the least concern is about the explainability of GNNs for networking problems. The study for the explainability and visualization of deep learning models has a long story and deep learning has been criticized for its ``black-box’’ property. The graph structure brings new challenges for the explainability problem. The development of post-processing techniques to explain the predictions made by GNNs has been made with some progresses, however, the explainability of GNNs in the networking domain has not yet been fully addressed~\cite{pope2019explainability, ying2019gnnexplainer}.

\section{Wireless Networks}
\label{sec:wireless}
In this section, we focus on the relevant studies in wireless network scenarios. For wireless networks, we refer to those transmitting information through wireless data connections without using a cable, including wireless local area network, cellular network, wireless ad hoc network, cognitive radio network, device-to-device (D2D) network, satellite network, vehicular network, etc. Some problems are ubiquitous in different formats of wireless networks, e.g., power control. We would first talk about these problems in general wireless network scenarios. Then we discuss the papers focusing on a specific wireless network scenario.

\subsection{General Wireless Network}
Compared with other deep learning models, GNNs have the advantage of handling the topology information, which may not be leveraged in previous studies with Euclidean deep learning models. In densely deployed wireless local area networks, the channel resource is limited. To increase the system throughput, the channels must be allocated efficiently. The features of the channel vectors with the topology information are extracted in~\cite{nakashima2020deep}, with the GCN model. Then a deep reinforcement learning is developed for channel allocation, which utilizes the features extracted by GCN. Topology information is also used in~\cite{zhang2019topology} for wireless network optimization. Combining a GE unit and a deep feed-forward network, a two-stage topology-aware framework is proposed and validated for the network flow optimization problem, which achieves a trade-off between computation time and inference performance.

Compared with the wired communication, wireless transmission may be imperfect with more errors. While GNNs may be applied in wireless networks, the transmission uncertainty would deteriorate the robustness of GNNs. This challenge is considered in~\cite{lee2021decentralized}, in which decentralized GNN binary classifiers are used for multiple problems, e.g., power control or wireless link scheduling. To handle this situation, re-transmission mechanisms are proposed to enhance the robustness of GNN classifiers, for both uncoded and coded wireless communication systems.

Power allocation or control is an important topic in the wireless network scenario, in which the devices connected to the network may be powered by batteries with a limited energy storage. The transmission in the free space may also interference with each other if the power is not properly controlled. To handle this problem, multiple GNN-based solutions are proposed~\cite{eisen2019large, eisen2020transferable, eisen2020optimal, chowdhury2021efficient, chowdhury2021unfolding, nikoloska2021fast, shen2019graph, shen2020graph, naderializadeh2020wireless}. In a series of studies~\cite{eisen2019large, eisen2020transferable, eisen2020optimal, nikoloska2021fast}, Random Edge Graph Neural Networks (REGNNs) are selected as the optimal solution for the power allocation and control optimization problem, with various system constraints. REGNNs outperform baselines with an essential permutation invariance property, which are desirable in networks of growing size. For the optimal power allocation in a single-hop ad hoc wireless network, an iterative weighted minimum mean squared error method named UWMMSE is proposed, in which GNNs are used to learn the model parameters~\cite{chowdhury2021efficient, chowdhury2021unfolding}. UWMMSE effectively reduces the computational complexity without harming the allocation performance, over the classic algorithm for power control. For solving the similar problem in an unsupervised approach, Interference Graph Convolutional Neural Network (IGCNet) is proposed and validated in~\cite{shen2019graph}, which is robust to imperfect Channel State Information (CSI). Beamforming is further considered in \cite{shen2020graph}, in which Message Passing Graph Neural Networks (MPGNNs) are proposed to solve both the power control and beamforming problems. Similarly, in an unsupervised approach to learn optimal power allocation decisions, a primal-dual counterfactual optimization approach is proposed in~\cite{naderializadeh2020wireless}, in which GNNs are used to handle the network topology.

To sum up, the papers in the general wireless network scenario are listed in Table~\ref{tab:wireless}. The target problem, proposed solution and the relevant GNN component(s) are also listed. The similar tabular format for the paper summary applies in the following sections.

\begin{table}[!htb]
\centering
\caption{List of the papers in the wireless network scenario.}
\label{tab:wireless}
\begin{tabular}{|p{3.5cm}|p{2cm}|p{3.3cm}|p{3cm}|}
\hline
Problem & Paper & Solution & GNN \\
\hline
Binary Classification & \cite{lee2021decentralized} & Decentralized GNN & GCN~\cite{kipf2017semi}, GIN~\cite{xu2019powerful} \\
\hline
Channel Allocation & \cite{nakashima2020deep} & DRL with GCN & ChebNet~\cite{defferrard2016convolutional} \\
\hline
Network Flow Optimization & \cite{zhang2019topology} & Two-stage Topology-aware ML Framework & MPNN~\cite{gilmer2017neural} \\
\hline
Power Allocation & \cite{eisen2019large, eisen2020transferable, eisen2020optimal} & REGNN & GNN~\cite{henaff2015deep} \\
\hline
Power Allocation & \cite{chowdhury2021efficient, chowdhury2021unfolding} & UWMMSE Method & GCN~\cite{kipf2017semi} \\
\hline
Power Control & \cite{nikoloska2021fast} & REGNN & GNN~\cite{henaff2015deep} \\
\hline
Power Control & \cite{shen2019graph} & IGCNet & GIN~\cite{xu2019powerful} \\
\hline
Power Control and Beamforming & \cite{shen2020graph} & MPGNNs & GIN~\cite{xu2019powerful}, GCN~\cite{kipf2017semi} \\
\hline
Power Control & \cite{naderializadeh2020wireless} & Unsupervised Primal-dual Counterfactual Optimization & GNN~\cite{henaff2015deep} \\
\hline
\end{tabular}
\end{table}

\subsection{Cellular Network}
Cellular networks are discussed separately in this part, not only because more than ten papers focus on this specific scenario, but also because the cellular network has a wide application. For example, there were 5.95 billion LTE subscriptions worldwide by the end of Q4 2020~\footnote{\url{https://gsacom.com/paper/lte-and-5g-subscribers-march-2021-q4/}}. While the growing trend may be affected by COVID-19, cellular networks are still one of the major approach for accessing the Internet.

Driven by the huge demand, the research in the cellular network scenario keeps increasing, including those leveraging graph-based deep learning models for some traditional communication problems, e.g., resource allocation, power control and traffic prediction. Driven by the ideas from SDN, some new problems also appear in the cellular network scenario, e.g., network slicing and virtual network embedding. Both types of problems have been investigated in the surveyed papers.

To fully utilize the network resources, multipath TCP is considered for 5G networks, which transfer packets over multiple paths concurrently. However, network heterogeneity in 5G networks makes the multipath routing problem become more complex for the existing routing algorithms to handle. A GNN-based multipath routing model is proposed as the solution in~\cite{zhu2020gclr}. The experiments under the SDN framework demonstrate that the GNN-based model can achieve a significant throughput improvement.

Traffic prediction is also considered in cellular networks, with GNN-based solutions being proposed in recent years~\cite{he2019graph, he2020graph, pan2021dc, sun2021mobile}. As a prediction problem, the temporal dependencies may be modeled by a recurrent neural network, e.g., Long Short Term Memory (LSTM) or GRU. Different attention mechanisms may also be incorporated. As an improvement over baselines, GNN is capable of modeling the spatial correlation between different nodes, e.g., a cell tower or an access point. Different structures have been explored in existing studies, e.g., GAT in~\cite{he2019graph, he2020graph}, GCN in~\cite{pan2021dc}, and GraphSAGE in~\cite{sun2021mobile}.

Energy consumption is another concern for 5G network, which is designed to enable a denser network with microcells, femtocells and picocells. To better control the transmission power, GNN-based power control solutions are proposed in~\cite{guo2021learning, hou2021user}. Heterogeneous GNNs (HetGNNs) with a novel parameter sharing scheme are proposed for power control in multi-user multi-cell networks~\cite{guo2021learning}. Take a step further, the joint optimization problem of user association and power control of the downlink is considered in~\cite{hou2021user}, in which an unsupervised GNN is used for power allocation and the Spectral Clustering algorithm is used for user association.

Green network management is proposed to improve the energy efficiency. A specific problem, the Idle Time Windows (ITWs) prediction, is considered in~\cite{fang2019idle}. To capture the spatio-temporal features, a novel Temporal Graph Convolutional Network (TGCN) is proposed for learning the network representation, which improves the prediction performance. Also for the denser cell sites, the Integrated Access and Backhaul (IAB) architecture defined by the 3rd Generation Partnership Project (3GPP) is used in~\cite{simsek2020iab}. The IAB topology design is formulated as a graph optimization problem and a combination of deep reinforcement learning and graph embedding is proposed for solving this problem efficiently.

The integration of satellite-terrestrial networks is proposed for the future 6G network. In this direction, a High Altitude Platform Station (HAPS) is a network node that operates in the stratosphere at an altitude around 20 km and is instrumental for providing communication services~\cite{kurt2021vision}. For HAPS, GAT is firstly utilized for channel estimation in~\cite{tekbiyik2021graph, tekbiyik2021channel}, and the proposed GAT estimator outperforms the traditional least square method in full-duplex channel estimation and is also robust to hardware imperfections and changes in small-scale fading characteristics.

As a softwarized concept, network slicing has been proposed for 5G network, using network virtualization to divide single network connection into multiple distinct virtual connections that provide services with different Quality-of-Service (QoS) requirements. However, the increasing network complexity is becoming a huge challenge for deploying network slicing. A scalable Digital Twin (DT) technology with GNN is developed in~\cite{wang2020graph} for mirroring the network behavior and predicting the end-to-end latency, which can also be applied in unseen network situations. Take a step further, GAT is incorporated into Deep Q Network (DQN) for designing an intelligent resource management strategy in~\cite{shao2021graph}, which is proven effective through simulations.

Virtual Network Embedding (VNE) is also a softwarized concept, which can be used for modeling the resource allocation of 5G network slices. Since the VNE problem is NP-hard, heuristic methods and deep learning models are both being proposed for this specific problem. Deep Reinforcement Learning (DRL) and GCN are combined for solving this problem~\cite{rkhami2020use, rkhami2021learn}, in which the episodic Markov Decision Process is solved by different GCN models.

To sum up, the papers in the cellular network scenario are listed in Table~\ref{tab:cellular}.
\begin{table}[!htb]
\centering
\caption{List of the papers in the cellular network scenario.}
\label{tab:cellular}
\begin{tabular}{|p{4cm}|p{1.5cm}|p{3.5cm}|p{2.5cm}|}
\hline
Problem & Paper & Solution & GNN \\
\hline
Channel Estimation & \cite{tekbiyik2021graph, tekbiyik2021channel} & GAT-based Estimator & GAT~\cite{velivckovic2018graph} \\
\hline
Idle Time Windows Prediction & \cite{fang2019idle} & TGCN & GCN~\cite{kipf2017semi} \\
\hline
Integrated Access and Backhaul Topology Design & \cite{simsek2020iab} & DRL with Graph Embedding & structure2vec~\cite{dai2016discriminative} \\
\hline
Network Modeling, Network Slicing & \cite{wang2020graph} & GNN-based Digital Twin & GraphSAGE~\cite{hamilton2017inductive} \\
\hline
Network Slicing & \cite{shao2021graph} & DQN with GAT & GAT~\cite{velivckovic2018graph} \\
\hline
Power Control & \cite{guo2021learning} & Heterogeneous GNNs & HetGAT~\cite{wang2019heterogeneous} \\
\hline
Routing & \cite{zhu2020gclr} & GCLR & MPNN~\cite{gilmer2017neural} \\
\hline
Traffic Prediction & \cite{sun2021mobile} & Graph-based TCN & GraphSAGE~\cite{hamilton2017inductive} \\
\hline
Traffic Prediction & \cite{he2019graph, he2020graph} & GASTN & S-RNN~\cite{jain2016structural} \\
\hline
Traffic Prediction & \cite{pan2021dc} & DC-STGCN & GCN~\cite{kipf2017semi} \\
\hline
User Association, Power Control & \cite{hou2021user} & Unsupervised Graph Model & GraphSAGE~\cite{hamilton2017inductive} \\
\hline
VNE & \cite{rkhami2020use, rkhami2021learn} & DRL with GCN & GCN~\cite{kipf2017semi} \\
\hline
\end{tabular}
\end{table}

\subsection{Other Wireless Networks}
In this part, we discuss the other formats of wireless networks, with their own challenges and solutions.

The first case is the cognitive radio network, which aims to increase the spectrum utilization by secondary users with an opportunistic use of the free spectrum that is not used by the primary users. In this scenario, the challenge is to improve the resource utilization, without degrading the quality of service (QoS) of primary users. To solve this challenge, a joint channel selection and power adaptation scheme is proposed in~\cite{zhao2020graph}, in which GCN is leveraged to extract the crucial interference features. Based on the estimated CSI, a DRL-based framework is further used to allocate spectrum resources efficiently.

The second case is the Device-to-Device (D2D) network, which uses the direct communication between two users or devices, without traversing the base station or router. Without deploying additional infrastructure, D2D network is promising for provide communication services with an ultra-low latency. However, there are still many challenges for this objective to happen. To minimize the content fetching delay in D2D network, the joint optimization of cooperative caching and fetching is considered in~\cite{yan2020cooperative} and a DRL-based algorithm is proposed. In the proposed algorithm, GAT is used for cooperative inter-agent coordination. For power control and beamforming in D2D network, an unsupervised learning-based framework is proposed in~\cite{zhang2021scalable}, in which heterogeneous graphs and GNNs are used for the characteristics of diversified link features and interference relations. Wireless link scheduling is also considered in a series of studies~\cite{lee2020wireless, lee2020graph, fu2020wireless}. Graph embedding based method is proposed in~\cite{lee2020wireless, lee2020graph}, in which the graph embedding process is based on the distances of both communication and interference links, without requiring the accurate CSI. The proposed method manages to reduce the computational complexity for the link scheduling problem significantly.

The third case is the Internet of Things (IoT) network, which is designed for connecting smart devices, e.g., smart meters, smart light bulbs, connected valves and pumps, etc. The application of IoT networks covers a wide range, e.g., smart factory, smart agriculture, smart city, etc. The wide application also arises a great number of challenges, e.g., resource utilization efficiency, battery limitation for computation and communication, and security concerns. Some of these challenges can be solved with graph-based methods. One example is the channel estimation problem considered in~\cite{tekbiyik2021graph}, in which Direct-to-satellite (DtS) communication is used for globally connected IoT networks and the high path loss must be considered. GAT is proposed as the solution and further used for the reconfigurable intelligent surfaces in the considered scenario. Another example is the network intrusion detection, which is drawing a growing attention in recent years. GraphSAGE is used in~\cite{lo2021graphsage} for using the edge features and classifying the network flows into benign and attack types. The new solution is proven more effective than the state-of-the-art methods on six benchmark datasets. SDN concepts are also applied in IoT networks and can be combined with graph-based solutions. NFV-enabled Service Function Chain (SFC) is considered in~\cite{liu2020dynamic}, in which the challenge is that SFCs should be dynamically and adaptively reconfigured in order to achieve a lower resource consumption and a higher revenue. This problem is formulated as a discrete-time Markov decision process and a deep Dyna-Q (DDQ) approach is proposed as the solution, in which GNNs are used for predicting available virtual network functions (VNFs).

The fourth case is the satellite network, in which the communication between satellites are considered. With the growing Low Earth Orbit (LEO) satellites launched by commercial companies, e.g., Starlink and OneWeb, satellite networks are drawing more attention, with a potential application in both IoT and future 6G networks. The traffic prediction problem in the satellite network is considered in~\cite{yang2020noval}, in which the spatial dependency of the network topology is captured by GCN and the temporal dependency is captured by GRU. The simulation using the satellite network traffic shows the combination with GCN improves the performance of the single GRU model.

The last case is the vehicular network, which aims to connect the vehicle nodes. Vehicular network has been proposed for autonomous driving in future smart cities, as an important infrastructure. One challenge is to improve the spectrum allocation efficiency. The vehicle-to-everything (V2X) network is considered in~\cite{he2020resource}, in which GNN is used to learn the low-dimensional feature and DRL is used to make spectrum allocation decisions. This kind of GNN-DRL combination has already been used in similar problems of other network types. Another challenge is to reduce the communication latency within vehicular networks, especially in the large-scale and fast-moving scenario. To model the communication latency between the vehicle and the infrastructure, a graph-based framework named SMART is proposed in~\cite{liu2021spatio}, in which GCN is combined with a deep Q-networks algorithm to capture the spatial and temporal patterns within a limited observation zone. Then the latency performance is re-constructed for the whole geographical area.

To sum up, the papers in other wireless network scenarios are listed in Table~\ref{tab:wireless_other}.

\begin{table}[!htb]
\centering
\caption{List of the papers specified in other wireless network scenarios.}
\label{tab:wireless_other}
\begin{tabular}{|p{3cm}|p{3cm}|p{1.5cm}|p{3cm}|p{2.5cm}|}
\hline
Scenario & Problem & Paper & Solution & GNN \\
\hline
Cognitive Radio Network & Resource Allocation & \cite{zhao2020graph} & DRL with GCN & GCN~\cite{kipf2017semi} \\
\hline
D2D Network & Cooperative Caching and Fetching & \cite{yan2020cooperative} & FDS-MARL & GAT~\cite{velivckovic2018graph} \\
\hline
D2D Network & Power Control and Beamforming & \cite{zhang2021scalable} & HIGNN & GN~\cite{battaglia2018relational} \\
\hline
D2D Network & Wireless Link Scheduling & \cite{lee2020graph, lee2020wireless} & Graph Embedding based Method & structure2vec~\cite{dai2016discriminative} \\
\hline
D2D Network & Wireless Link Scheduling & \cite{fu2020wireless} & Graph Embedding based Method & structure2vec~\cite{dai2016discriminative} \\
\hline
IoT Network & Intrusion Detection & \cite{lo2021graphsage} & E-GraphSAGE & GraphSAGE~\cite{hamilton2017inductive} \\
\hline
IoT Network & Service Function Chain Dynamic Reconfiguration & \cite{liu2020dynamic} & Deep Dyna-Q Approach & GNN~\cite{scarselli2008graph} \\
\hline
Satellite Network & Traffic Prediction & \cite{yang2020noval} & GCN-GRU & GCN~\cite{kipf2017semi} \\
\hline
Vehicular Network & Communication Latency Modeling & \cite{liu2021spatio} & SMART Framework & GCN~\cite{kipf2017semi} \\
\hline
Vehicular Network & Spectrum Allocation & \cite{he2020resource} & DQN with GNN & GNN~\cite{scarselli2008graph} \\
\hline
\end{tabular}
\end{table}

\section{Wired Networks}
\label{sec:wired}
For wired networks, we mainly refer to the computer networks that are connected with cables, such as laptop or desktop computers. A typical example is the Ethernet network. In this section, we first discuss the graph-based studies in the wired network scenario from five aspects, namely, network modeling, network configuration, network prediction, network management, and network security. Then three special cases are further discussed, i.e., blockchain platform, data center network, and optical network.

GNNs are suitable for network modeling as the computer networks are often modeled as graphs. With the growing trend of contemporary Internet, it becomes more and more challenging to understand the overall network topology, the architecture and different elements of the networks, and their configurations. To solve this challenge, GNNs are proposed for network modeling. They are not only used to reconstruct the existing networks, but also used to model the non-existing networks, in order to provide an estimation of the unseen cases for network operators to make better network deployment decisions in the future. By modeling networks, the estimation of different end-to-end metrics are concerned in surveyed studies, given the input network topology, routing scheme and traffic matrices of the network, in a supervised~\cite{suarez2019challenging, badia2019towards, ferriol2020applying, geyer2019deepcomnet} or semi-supervised~\cite{suzuki2020estimating} way. Delay and jitter are considered in~\cite{suarez2019challenging, badia2019towards, ferriol2020applying, suzuki2020estimating}, while the throughput of TCP flows and the end-to-end latency of UDP flows are considered in~\cite{geyer2019deepcomnet}. Different GNNs are used for the network modeling purpose, including GGS-NN in~\cite{geyer2019deepcomnet}, MPNN in~\cite{suarez2019challenging, badia2019towards}, GN and GNN in~\cite{ferriol2020applying}, and GCN in~\cite{suzuki2020estimating}. GNNs are also used for network calculus analysis in~\cite{geyer2019deeptma, geyer2020graph, geyer2020robustness}.

Based on the modeling ability of GNNs, they are further proposed for network configuration feasibility analysis or decision. Based on the prediction of ensemble GNN model, different network configurations are evaluated in~\cite{mai2021improvements}, bound to the deadline constraints. Border Gateway Protocol (BGP) configuration synthesis is considered in~\cite{bahnasy2020deepbgp}, which is the standard inter-domain routing protocol to exchange reachability information among Wide Area Networks (WANs). GNN is adopted to represent the network topology with partial network configuration in a system named DeepBGP, which is further validated for both Huawei and Cisco devices while fulfilling operator requirements. Another relevant study is to use GNN for Multiprotocol Label Switching (MPLS) configuration analysis. A GNN-based solution named DeepMPLS is proposed in~\cite{geyer2019deepmpls} to speed up the analysis of network properties as well as to suggest configuration changes in case a network property is not satisfied. The GNN-based solution manages to achieve low execution times and high accuracies in real-world network topologies.

GNNs can also be used for network prediction, e.g., delay prediction~\cite{rusek2018message} and traffic prediction~\cite{zhao2020spatiotemporal, yang2020mstnn, mallick2020dynamic}. The better prediction is the basis of proactive management. A case study of delay prediction in queuing networks is conducted in~\cite{rusek2018message}, which uses MPNN for topology representation and network operation. Several studies are concerned about data-driven traffic prediction, based on the real-world network traffic data and GNN-based solutions. A framework named Spatio-temporal Graph Convolutional Recurrent Network (SGCRN) is proposed in~\cite{zhao2020spatiotemporal}, which combines GCN and GRU and is validated on the network traffic data from four real IP backbone networks. Another framework named Multi-scale Spatial-temporal Graph Neural Network (MSTNN) is proposed for Origin-Destination Traffic Prediction (ODTP) and two real-world datasets are used for evaluation~\cite{yang2020mstnn}. Inspired by the prediction model DCRNN~\cite{li2018dcrnn_traffic} developed for road traffic, a nonautoregressive graph-based neural network is used in~\cite{mallick2020dynamic} for network traffic prediction and evaluated on the U.S. Department of Energy's dedicated science network.

Network prediction results can be used further for network operation optimization and management~\cite{otoshi2015traffic}, e.g., traffic engineering, load balancing, routing, etc. For the time point of preparing this survey, routing is considered with graph-based deep learning models~\cite{geyer2018learning, xiao2020neural}. Instead of using reinforcement learning, a novel semi-supervised architecture named Graph-Query Neural Network is proposed in~\cite{geyer2018learning} for shortest path and max-min routing. Another graph-based framework named NGR is proposed in~\cite{xiao2020neural} for shortest-path routing and load balancing. These graph-based routing solutions are validated with use-cases and show high accuracies and resilience to packet loss.

Last but not the least, graph-based deep learning solutions are used for network security problems in computer networks~\cite{zhou2020auto, cheng2021discovering}. Automatic detection for Botnets, which is the source of DDoS attacks and spam, is considered in~\cite{zhou2020auto}. GNN is used to detect the patterns hidden in the botnet connections and is proven more effective than non-learning methods. Their dataset is also made available for future studies. In another study, intrusion detection is considered~\cite{cheng2021discovering}. A GCN-based framework named Alert-GCN is proposed to solve the intrusion alert problem as a node classification task. The alert graph is built with the alert information from farther neighbors, which is used as the input for the GCN module. The experiments demonstrate that Alert-GCN outperforms traditional classification models in correlating alerts.

To sum up, the papers in the wired network scenario are listed in Table~\ref{tab:wired}.
\begin{table}[!htb]
\centering
\caption{List of the papers in the wired network scenario.}
\label{tab:wired}
\begin{tabular}{|p{3.5cm}|p{2cm}|p{3.3cm}|p{3cm}|}
\hline
Problem & Paper & Solution & GNN \\
\hline
BGP Configuration Synthesis & \cite{bahnasy2020deepbgp} & DeepBGP & GraphSAGE~\cite{hamilton2017inductive}, GNN~\cite{scarselli2008graph} \\
\hline
Botnet Detection & \cite{zhou2020auto} & GNN Approach & GCN~\cite{kipf2017semi} \\
\hline
Communication Delay Estimation & \cite{suzuki2020estimating} & GNNs with Semi-supervised Learning & GCN~\cite{kipf2017semi} \\
\hline
Delay Prediction & \cite{rusek2018message} & Message-passing Neural Networks & MPNN~\cite{gilmer2017neural} \\
\hline
Intrusion Detection & \cite{cheng2021discovering} & Alert-GCN & GCN~\cite{kipf2017semi} \\
\hline
MPLS Configuration Analysis & \cite{geyer2019deepmpls} & DeepMPLS & GNN~\cite{scarselli2008graph} \\
\hline
Network Calculus Analysis & \cite{geyer2019deeptma, geyer2020graph, geyer2020robustness} & DL-assisted Tandem Matching Analysis & GNN~\cite{scarselli2008graph} \\
\hline
Network Configuration Feasibility & \cite{mai2021improvements} & Ensemble GNN Model & GN~\cite{battaglia2018relational} \\
\hline
Network Modeling & \cite{suarez2019challenging} & RouteNet & MPNN~\cite{gilmer2017neural} \\
\hline
Network Modeling & \cite{badia2019towards} & Extended RouteNet & MPNN~\cite{gilmer2017neural} \\
\hline
Network Modeling & \cite{ferriol2020applying} & Graph-based DL & GN~\cite{battaglia2018relational}, GNN~\cite{scarselli2008graph} \\
\hline
Network Modeling & \cite{geyer2019deepcomnet} & DeepComNet & GGS-NN~\cite{li2016gated} \\
\hline
Routing & \cite{geyer2018learning} & Graph-Query Neural Network & GNN~\cite{scarselli2008graph} \\
\hline
Routing and Load Balancing & \cite{xiao2020neural} & DL-based Distributed Routing & GNN~\cite{scarselli2008graph} \\
\hline
Traffic Prediction & \cite{zhao2020spatiotemporal} & SGCRN & GCN~\cite{kipf2017semi}\\
\hline
Traffic Prediction & \cite{yang2020mstnn} & MSTNN & GAT~\cite{velivckovic2018graph} \\
\hline
Traffic Prediction & \cite{mallick2020dynamic} & Nonautoregressive Graph-based Neural Network & DCRNN~\cite{li2018dcrnn_traffic} \\
\hline
\end{tabular}
\end{table}

Other than the general computer network case, three specific network cases are discussed with graph-based methods.

The first case is the blockchain platform, which is well-known by the public thanks to Bitcoin, the most famous cryptocurrency. Generally speaking, the blockchain is a chain of blocks that store information with digital signatures in a decentralized and distributed network, which has a wide range of applications other than digital cryptocurrencies, e.g., financial and social services, risk management, healthcare facilities, etc~\cite{monrat2019survey}. A specific task of encrypted traffic classification is considered in~\cite{shen2021accurate} for Decentralized Applications (DApps). A GNN-based DApp fingerprinting method named GraphDApp is proposed for this task and a novel graph structure named Traffic Interaction Graph (TIG) is constructed as the representation of encrypted DApp flows as well as the input for GNNs. Real-world traffic datasets from 1,300 DApps with more than 169,000 flows are used for experiments, of which the result shows that GraphDApp is superior to the other state-of-the-art methods in terms of classification accuracy.

The second case is the data center network, which connects all data centers to share data or computation abilities. Nowadays, data centers are heavily used for cloud services. In such circumstances, traffic engineering is becoming more and more important for the data center network to avoid traffic congestion and improve routing efficiency. However, this task is still challenging, especially when the network topology changes. In a recent study~\cite{li2020traffic}, the generalization ability of GNNs is used for predicting Flow Completion Time (FCT) and a GNN-based optimizer is further designed for flow routing, flow scheduling and topology management. The experiments demonstrate both the high inference accuracy and the FCT reduction ability of GNNs.

The last case is the optical network, which uses light signals, instead of electronic ones, to send information between two or more points. There are many unique problems when light signals are used for communication, e.g., wavelength assignment. The optimal resource allocation in a special network type, i.e., Free Space Optical (FSO) fronthaul network, is considered in~\cite{gao2020resource} and GNNs are used for evaluating and choosing the resource allocation policy. The routing optimization for an Optical Transport Network (OTN) scenario is considered in~\cite{almasan2019deep} and the learning and generalization capabilities of GNNs are combined with DRL for routing in unseen network typologies. Similar to cellular and computer networks, traffic prediction is also considered in the optical network scenario~\cite{gui2020optical}, with the solution combined by GCN and GRU.

To sum up, the papers in other wired network scenarios are listed in Table~\ref{tab:wired_other}.
\begin{table}[!htb]
\centering
\caption{List of the papers specified in other wired network scenarios.}
\label{tab:wired_other}
\begin{tabular}{|p{3cm}|p{3cm}|p{1.5cm}|p{3cm}|p{2.5cm}|}
\hline
Scenario & Problem & Paper & Solution & GNN \\
\hline
Blockchain Platform & Encrypted Traffic Classification & \cite{shen2021accurate} & GNN-based DApps Fingerprinting & GIN~\cite{xu2019powerful} \\
\hline
Data Center Network & Traffic Optimization & \cite{li2020traffic} & GNN-based Optimizer & GN~\cite{battaglia2018relational} \\
\hline
Optical Network & Resource Allocation & \cite{gao2020resource} & GNN & GNN~\cite{henaff2015deep} \\
\hline
Optical Network & Routing & \cite{almasan2019deep} & DRL with GNN & MPNN~\cite{gilmer2017neural} \\
\hline
Optical Network & Traffic Prediction & \cite{gui2020optical} & GCN-GRU & GCN~\cite{kipf2017semi} \\
\hline
\end{tabular}
\end{table}

\section{Software Defined Networks}
\label{sec:sdn}
SDN emerges as the most promising solution for bringing a revolution in how networks are built. Based on the white paper released by the Open Networking Foundation (ONF), the explosion of mobile devices and content, server virtualization, and advent of cloud services are among the trends driving the networking industry to reexamine traditional network architectures~\footnote{\url{https://opennetworking.org/sdn-resources/whitepapers/software-defined-networking-the-new-norm-for-networks/}}. While SDN was proposed back to 1996, its concept has gone through a lot of changes ever since then. Based on the a widely used definition in~\cite{sezer2013we}, in the SDN architecture, the control and data planes are decoupled, network intelligence and state are logically centralized, and the underlying network infrastructure is abstracted from the applications.

The central control ability of SDN becomes the basis of network optimization in many scenarios and arises several problems which are in the scope of graph-based deep learning methods. Based on the surveyed studies in this paper, there is a growing trend of using GNNs with SDN, or the SDN concept in specific network scenarios. The benefits of this combination are two-folds. For GNNs, SDN provides the ability of measuring network performance, which is used as the data for training GNNs. For SDN, GNNs act as the best option of using the network topology information in modeling and optimizing the networks. In recent years, many graph-based solutions are proposed for various problems with the SDN concept.

Based on topology, routing, and input traffic, MPNN-based network models are proven to produce accurate estimates of the per-source/destination per-packet delay distribution and loss, with a worst Mean Relative Error (MRE) of 15.4\%, and the estimation can be further used for efficient routing optimization and network planning~\cite{rusek2019unveiling, rusek2020routenet}. The decoupling of the control plane and data plane gives more computing power for routing optimization. Based on this observation, an intelligent routing strategy based on graph-aware neural networks is designed in~\cite{zhuang2019toward}, in which a novel graph-aware convolution structure is constructed to learn topological information efficiently. In another study for routing optimization, a GN-based solution is proposed for maximum bandwidth utilization, which achieves a satisfactory accuracy and a prediction time 150 times faster than Genetic Algorithm (GA)~\cite{sawada2020network}.

In SDN, network virtualization is a powerful way to efficient utilize the network infrastructure. Virtual Network Functions (VNFs) are virtualized network services running on physical resources. How to map VNFs into shared substrate networks has become a challenging problem in SDN, known as Virtual Network Embedding (VNE) or VNF placement, which is already proven to be NP-hard. To efficiently solve this problem, a bunch of heuristic algorithms are proposed in the literature. Recently, graph-based models have also been used for this problem~\cite{mijumbi2016connectionist, mijumbi2017topology, kim2020graph1, kim2021graph, habibi2020accelerating, kim2020graph, sun2020combining}, which can get near-optimal solutions in a short time. To predict future resource requirements for VNFs, a GNN-based algorithm using the VNF forwarding graph topology information is proposed in~\cite{mijumbi2016connectionist, mijumbi2017topology}. Deployed in a virtualized IP multimedia subsystem and tested with real VoIP traffic traces, the new algorithm achieves an average prediction accuracy of 90\% and improves the call setup latency by over 29\%, compared with the case without using GNNs. A parallelizable VNE solution based on spatial GNNs is proposed for accelerating the embedding process in~\cite{habibi2020accelerating}, which improves the revenue-to-cost ratio by about 18\%, compared to other simulated algorithms. Similarly, GNN-based algorithms are proposed for VNF resource prediction and management in a series of studies~\cite{kim2020graph1, kim2021graph, kim2020graph}. On another aspect, DRL is often combined with GNNs for automatic virtual network embedding~\cite{jalodia2019deep, yan2020automatic, sun2020deepmigration, sun2020efficient}. Asynchronous DRL enhanced GNN is proposed in~\cite{jalodia2019deep} for topology-aware VNF resource prediction in dynamic environments. An efficient algorithm combining DRL with GCN is proposed in~\cite{yan2020automatic}, with up to 39.6\% and 70.6\% improvement on acceptance ratio and average revenue, compared with the existing state-of-the-art solutions. A more specific problem, i.e., traffic flow migration among different network function instances, is considered in~\cite{sun2020deepmigration, sun2020efficient}, in which GNN is used for migration latency modeling and DRL is used for deploying dynamic and effective flow migration policies.

Last but not the least, Service Function Chaining (SFC) is considered in several studies~\cite{heo2020reinforcement, heo2020graph, rafiq2020service, wang2021drl}. SFC uses SDN's programmability to create a service chain of connected virtual network services, resulting in a service function path that provides an end-to-end chain and traffic steering through them. Graph-structured properties of network topology can be extracted by GNNs, which outperforms DNNs for SFC~\cite{heo2020graph, rafiq2020service}. However, most of the existing studies for SFC use a supervised learning approach, which may not be suitable for dynamic VNF resources, various requests, and changes of topologies. To solve this problem, DRL is applied for training models on various network topologies with unlabeled data in~\cite{heo2020reinforcement} and achieves remarkable flexibility in new topologies without re-designing and re-training, while preserving a similar level of performance compared to the supervised learning method. DRL is also used for adaptive SFC placement to maximize the long-term average revenue~\cite{wang2021drl}.

To sum up, the papers in the SDN scenario are listed in Table~\ref{tab:sdn}.
\begin{table}[!htb]
\centering
\caption{List of the papers in the SDN scenario.}
\label{tab:sdn}
\begin{tabular}{|p{3.5cm}|p{2cm}|p{3.3cm}|p{3cm}|}
\hline
Problem & Paper & Solution & GNN \\
\hline
Network Modeling & \cite{rusek2019unveiling, rusek2020routenet} & RouteNet & MPNN~\cite{gilmer2017neural} \\
\hline
Routing & \cite{zhuang2019toward} & Revised Graph-aware Neural Networks & A Novel Graph-aware Convolution Structure \\
\hline
Routing Optimization, Bandwidth Utilization Maximization & \cite{sawada2020network} & GN-based Model & GN~\cite{battaglia2018relational} \\
\hline
SFC & \cite{heo2020reinforcement} & DRL with GNN & GNN~\cite{scarselli2008graph} \\
\hline
SFC & \cite{heo2020graph} & GNN-based SFC & GCN~\cite{kipf2017semi} \\
\hline
SFC Deployment, Traffic Steering & \cite{rafiq2020service} & Knowledge-Defined Networking System with GNN & GNN~\cite{scarselli2008graph} \\
\hline
SFC Placement & \cite{wang2021drl} & DRL-SFCP & GCN~\cite{kipf2017semi} \\
\hline
Traffic Flow Migration in NFV & \cite{sun2020deepmigration, sun2020efficient} & DRL with GNN & GN~\cite{battaglia2018relational} \\
\hline
VNE & \cite{habibi2020accelerating} & GraphViNE Solution & GraphSAGE~\cite{hamilton2017inductive}, GE~\cite{pan2019learning} \\
\hline
VNE & \cite{yan2020automatic} & DRL with GCN & GCN~\cite{kipf2017semi} \\
\hline
VNF Deployment Prediction & \cite{kim2020graph1, kim2021graph} & GNN-based Algorithm & GNN~\cite{scarselli2008graph} \\
\hline
VNF Management & \cite{kim2020graph} & GNN-based Algorithm & GNN~\cite{scarselli2008graph} \\
\hline
VNF Placement & \cite{sun2020combining} & DRL with GNN & GN~\cite{battaglia2018relational} \\
\hline
VNF Resource Prediction & \cite{jalodia2019deep} & Asynchronous DRL enhanced GNN & GNN~\cite{scarselli2008graph} \\
\hline
VNF Resource Prediction & \cite{mijumbi2016connectionist, mijumbi2017topology} & GNN-based Algorithm & GNN~\cite{scarselli2008graph} \\
\hline
\end{tabular}
\end{table}

\section{Future Directions}
\label{sec:direction}
In this section, we discuss some future directions for graph-based deep learning in communication networks. Even though different network scenarios and applications are already covered in this survey, there are still many open research opportunities for this topic.

The first research direction is the combination of GNNs and other artificial intelligence techniques. Some examples are already seen in this survey, e.g., the combination of GNN and GRU for traffic prediction~\cite{yang2020noval, gui2020optical}, the combination of GNN and DRL for resource allocation~\cite{zhao2020graph}, routing~\cite{almasan2019deep}, and VNE~\cite{yan2020automatic}. The advantages of GNNs include its learning ability for topological dependencies and the generalization capability for unseen network typologies, but GNN is not a panacea. For example, for some cases which is lack of training data or is too expensive to collect real data, Generative Adversarial Nets (GANs)~\cite{goodfellow2014generative} is a possible solution. Even though GANs have been widely used in other fields, e.g., image and video, the combination of GANs and GNNs~\cite{wang2018graphgan} has not been applied for communication networks, at least in the scope of this survey. Another example is the Automated Machine Learning (AutoML) technique~\cite{he2021automl}, which can be used for optimizing the GNN parameters automatically.

Another research direction is to apply graph-based deep learning on larger networks. In most of the surveyed studies, the network topology is small, e.g., less than 100 nodes, compared with contemporary networks. However, the modeling of larger networks would require huge computation requirements. Graph partitioning and parallel computing infrastructures are two possible solutions for this problem. A larger network may be decomposed into smaller ones that is within the computing capacity. However, the optimal divide-and-conquer approach remains unknown and may vary in different network scenarios. Another concern is that whether it is worthy of achieving narrow performance margins in the cost of the increased computation burden caused by graph-based models, compared with traditional methods.

Finally, we believe this is still an early stage of the research about graph-based deep learning for communication networks. There are many opportunities of applying novel GNNs in traditional networking problems in a wider range of network scenarios, especially those who get little or no attention for now. The studies covered in this survey are only the beginning of this exciting research area. And we would keep track of this area and update the progress and new publications in the public Github repository.

\section{Conclusion}
\label{sec:conclusion}
In this paper, a survey is presented for the application of graph-based deep learning in communication networks. The relevant studies are organized in three network scenarios, namely, wireless networks, wired networks, and software defined networks. For each study, the problem and GNN-based solution are listed in this survey. Future directions are further pointed out for the follow-up research. We hope this survey could be the milestone of summarizing the latest progresses and a reference manual for new-comers in this emerging research topic.

\nocite{*}
\bibliography{mybibfile}

\end{document}